17[th] Century Photometric Data in the Form of Johannes Hevelius's Telescopic Measurements of the Apparent Diameters of Stars

Christopher M. Graney
Jefferson Community & Technical College
1000 Community College Drive
Louisville, Kentucky (USA)
christopher.graney@kctcs.edu

ABSTRACT

Johannes Hevelius's 1662 *Mercurius in Sole Visus Gedani* contains a table of magnitudes and apparent telescopic diameters of nineteen stars.  The data conform to a simple model, suggesting that Hevelius produced what is essentially a table of surprisingly precise photometric data.

INTRODUCTION

In 1662 Johannes Hevelius published in his *Mercurius in Sole Visus Gedani* a table of observations of stars that included, among other things, their magnitudes and apparent telescopic diameters (Figure 1).  Hevelius's data are consistent with a simple mathematical model published in this journal that explains, via circular aperture diffraction and a detection threshold, the apparent stellar diameters reported by early telescopic astronomers such as Galileo Galilei (Graney 2007; Graney and Sipes 2009).  Hevelius's data provide support for the model and context for understanding early telescopic observations.  But Hevelius's data also suggest that telescopic astronomers in the 17$^{th}$ century could obtain, in their measurements of apparent diameters of stars, what is essentially photometric data of some precision.  Besides being an interesting point of astronomical history, Hevelius's work suggests that useful historical photometric data may be available to researchers who are willing to look for it -- a task that is becoming easier with the passage of time, thanks to the growing collection of historical material available electronically.

THE HISTORY BOOKS ON EARLY PHOTOMETRY

Historians of astronomy generally report that early telescopic astronomers did not take much interest in the stars -- that for almost two centuries after the advent of the telescope the stars remained primarily reference points in the sky against which to measure the positions of solar system objects; with the exception of William Herschel, few astronomers studied stars carefully; what interest existed concerning stars was focused on variable stars (Pannekoek 1961, pp. 311-312; Mason 1962, pp. 298-300; Hoskin 1997, pp. 198-201). According to the noted historian of astronomy Michael Hoskin (1997), even the efforts to study variable stars "seemed to be leading astronomy nowhere", and were falling out of favor by the end of the 17th century (p. 201). Hoskin goes on to say that

> Part of the problem lay in the lack of a sufficiently delicate technique for monitoring the apparent brightness of a star. Stars were simply grouped according to the crude classification inherited from Antiquity, whereby the brightest stars were first magnitude and the faintest, sixth. The mid-nineteenth century would see the invention of new instruments to give an objective measure of the brightness of stars, and a new definition of magnitude. But before then, in the closing years of the eighteenth century, astronomers were at last provided with a simple method of determining whether a star had in fact altered in brightness [this being William Herschel's "Catalogues of the Comparative Brightnesses of Stars"; pp. 201-202].

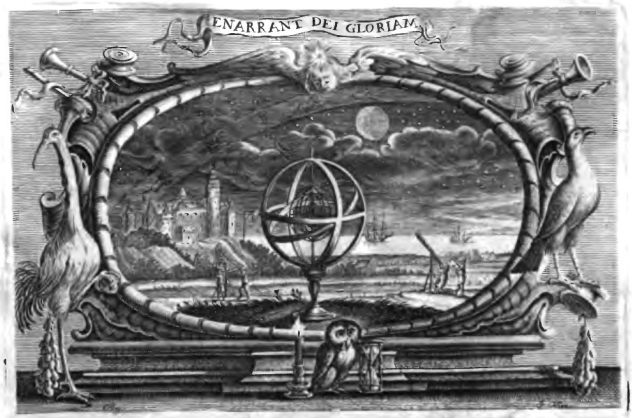

**FIGURE 1** -- Johannes Hevelius's 1662 *Mercurius in Sole Visus Gedani*.

J. B. Hearnshaw (1996) echoes this sentiment in *The Measurement of Starlight: Two Centuries of Astronomical Photometry,* his book on the history of photometry:

> The invention of the telescope...represents one of the great moments in the history of astronomy. Yet the telescope had relatively little beneficial effect on the quality of stellar magnitudes in the seventeenth and eighteenth centuries [p. 12].

More recently Richard Miles (2007) has said the same thing in his discussion of photometry's history:

> Though the arrival of the telescope marked the dawning of a new age, a further 170 years or so passed before someone properly applied scientific principles to visual magnitude estimates, when William Herschel (1738-1822) produced the first reliable naked-eye estimates of stars [p. 173].

In other words, until Herschel created a method of comparative photometry, no reliable method of photometry existed. On the subject of photometry Hoskin (1997) goes on to say that

> Since Classical Antiquity stars had been assigned a magnitude, the brightest in the sky being of first magnitude and the faintest visible on a clear night of sixth; the five intervals between were estimated. When the first telescopes revealed many still fainter stars, the scale was extended by little more than guesswork; a star classed by one astronomer as eighth magnitude might be described by another observer as eleventh [pp. 296-297].

Hearnshaw is of the same opinion:

> Although fainter stars became accessible [owing to the telescope], there were no guidelines concerning what magnitudes should be assigned to them, so that the scales of different observers diverged widely [p. 12-13].

Hoskin, as seen in the first quote from him above, sets the dawn of practical photometry in the 19th century. This view prevails in a variety of history of astronomy sources published over a long span of time (Grant 1852, p. 541; King 1955, pp. 295-297; Miczaika and Sinton 1961, pp. 154-157; Pannekoek 1961, pp. 385-387; Hearnshaw 1996, p. 105; Hoskin 1997, pp. 201, 297; Miles 2007, pp. 174-175).

HEVELIUS'S PHOTOMETRIC DATA

Recent work by this author and others has shown that, from the very beginning of telescopic astronomy, telescopic astronomers took interest in and made detailed observations of the stars. In particular, astronomers such as Simon Marius and Galileo Galilei report that stars seen through a telescope have noticeable disks -- disks which are larger for brighter stars and smaller for fainter stars (Ondra 2004, Siebert 2005, Graney 2007).

This author has proposed that these disks, which are of course spurious in nature, are simply the visible central maxima of Airy patterns formed via diffraction.  All stars seen with a given telescope at a given wavelength have the same Airy Disk Radius.  However, the eye cannot see indefinitely low levels of intensity.  Thus while the diffraction pattern for a circular aperture consists of a central maximum and an infinite number of rings of declining intensity, usually only a few diffraction rings are visible to the eye (Born and Wolf 1999, p. 442).  The rest fall below the eye's threshold of detection.  If the peak intensity of the pattern is low enough, no rings will be visible (for none cross the detection threshold) and the radius of the visible central maximum will be significantly smaller than the Airy Disk Radius.  For progressively lower peak intensities, the radius of the visible central maximum will decrease (Figure 2).  The result is that, for a telescope whose aperture is sufficiently small versus the magnification used, stars will appear as disks, with brighter stars having larger disks than fainter stars.  The relationship between a star's magnitude and the diameter of its image seen through a telescope will appear roughly linear to an observer who is measuring these quantities visually.  Thus when Galileo states in his *Dialogue Concerning the Two Chief World Systems* that stars of magnitude 1 measure 5" in diameter while stars of magnitude 6 measure 5/6" in diameter (Galileo 1967, p. 359), he is reporting results consistent with what we would expect from telescopes of the size he used (Graney 2007, pp. 446-448; Graney and Sipes 2009 pp. 98-103).

The 1662 Hevelius data support this view.  While Galileo's statement in the *Dialogue* implies that star image diameters decrease linearly with magnitude, Hevelius explicitly shows them to decrease linearly with magnitude (Table 1; Figure 3). However, the more interesting results are found from comparing Hevelius's star image diameters against modern magnitude measurements from the SIMBAD database. Hevelius's data conform remarkably well to the results of the Airy pattern with threshold (APT) model (Table 2; Figure 4). The model's output is in turn consistent with the sort of instrument Hevelius might use (Figure 5). Besides providing support for the APT model, this suggests that Hevelius's measurements, essentially a form of photometry, were capable of detecting small differences in stellar magnitudes. Hevelius shows us that even in the $17^{th}$ century it was possible to obtain photometric data that was surprisingly precise, given what historians have had to say about the subject -- a significant improvement over estimates or guesswork. Hevelius even provides a more precise alternative to the magnitude scale, an alternative in which Sirius measures 60 units, a sixth-magnitude star measures 18 units, and stars of the same magnitude class differ by as little as 2 units (see Table 1).

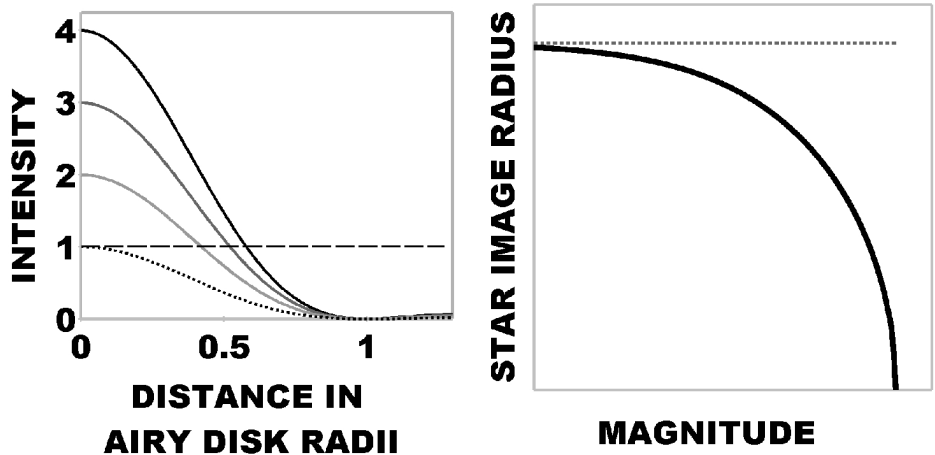

FIGURE 2 -- The Airy pattern with threshold (APT) model.  Left:  The author's plots of the classic Airy pattern formed by diffraction of light through a circular aperture.  The dashed horizontal line represents the detection threshold of the eye -- those parts of a star's Airy pattern that fall below the threshold are not seen.  Curve 1 (dotted) is for a star whose peak intensity falls just below the threshold; the star is not visible to the eye.  Brighter stars whose peak intensities are progressively higher will be visible to the eye, and will appear progressively larger.  Curves 2, 3, and 4 represent stars with peak intensities 2, 3, and 4 times that of star 1.  All the stars have the same Airy Disk Radius, but the visible radius of each star is defined by where its Airy pattern curve drops below the threshold.  Thus star 3 appears larger than star 2, and star 4 appears larger than star 3.  Right:  A general plot of star image radius vs. magnitude for the APT model.  The curve goes to zero for a star whose magnitude defines the threshold, such as star 1 in the left-hand plot.  Brighter stars have increasing size, approaching the Airy Disk Radius (horizontal dotted line) for progressively lower magnitudes.  This was sorted out in the 19$^{th}$ century by astronomers trying to reconcile the Airy Disk Radius concept to their visual observations that star image sizes decreased with magnitude (Hunt 1879).

CONCLUSIONS

Hevelius's data leads to two conclusions. The first conclusion is that the work of Galileo concerning the apparent telescopic sizes of stars, exemplified by his statement in his *Dialogue* that stars of magnitude 1 measure 5" in diameter while stars of magnitude 6 measure 5/6" in diameter, and discussed in this journal to great extent in previous papers

Rationes Stellarum Fixarum ad Terram, Solem or-
bitamq; magnam Terræ.

| Nomina Fixarum. | Magnitudo. | Quibus circellis fuerint æquales Part. | Diameter Fixarum apparēs. se. ter. | Vera Semid. in mill. juxt. Tych. & S.T. juxt Auctorē. | Ratio soliditatis stellæ ad soliditatem Terræ. | Ratio soliditatis orbis magni ad Stellas. | Ratio Stellarum ad Solem. |
|---|---|---|---|---|---|---|---|
| Sirius | 1 | 60 | 6 21 | 82 T. | 1000 Min. | | 125000 Min. |
|  |  |  |  | 918 H. | 773620632 Maj. | 216 Maj. | 72653 Maj. |
| Lucida Lyræ | 1 | 58 | 6 16 | 82 T. | 1000 Min. | | 125000 Min. |
|  |  |  |  | 912 H. | 758550528 Maj. | 216 Maj. | 71239 Maj. |
| Regel Orionis | 1 | 56 | 6 3 | 79 T. | 1331 Min. | | 166375 Min. |
|  |  |  |  | 876 H. | 672221376 Maj. | 216 Maj. | 63131 Maj. |
| Capella | 1 | 56 | 6 3 | 79 T. | 1331 Min. | | 166375 Min. |
|  |  |  |  | 876 H. | 672221376 Maj. | 216 Maj. | 63131 Maj. |
| Arcturus | 1 | 56 | 6 3 | 79 T. | 1331 Min. | | 166375 Min. |
|  |  |  |  | 876 H. | 672221376 Maj. | 216 Maj. | 63131 Maj. |
| Palilicium | 1 | 52 | 5 37 | 74 T. | 1728 Min. | | 167375 Min. |
|  |  |  |  | 816 H. | 541343375 Maj. | 216 Maj. | 50838 Maj. |
| Spica | 1 | 50 | 5 24 | 71 T. | 1728 Min. | | 216000 Min. |
|  |  |  |  | 786 H. | 485587656 Maj. | 343 Maj. | 45603 Maj. |
| Regulus Leonis | 1 | 48 | 5 11 | 66 T. | 2197 Min. | | 274625 Min. |
|  |  |  |  | 726 H. | 382657176 Maj. | 343 Maj. | 35937 Maj. |

TABLE 1a -- Hevelius's data from *Mercurius in Sole Visus Gedani* (1662, p. 94). The second column is magnitude and the third is diameter measured in seconds and thirds (1/60") of arc.

| | | | | | | | | |
|---|---|---|---|---|---|---|---|---|
| Prima Caudæ ursæ Majoris. | 2 | 46 | 4 58 | 65 T. | 2197 Min. | | 274625 Min. |
| | | | | 720 H. | 373248000 Maj. | 343 Maj. | 35052 Maj. |
| Procyon | 2 | 46 | 4 58 | 65 T. | 2197 Min. | | 274625 Min. |
| | | | | 720 H. | 373248000 Maj. | 343 Maj. | 35052 Maj. |
| Lucida Coronæ | 2 | 42 | 4 32 | 59 T. | 2744 Min. | | 343000 Min. |
| | | | | 660 H. | 287496000 Maj. | 512 Maj. | 27000 Maj. |
| Caput Serpentarii | 2 | 42 | 4 32 | 59 T. | 2744 Min. | | 343000 Min. |
| | | | | 660 H. | 287496000 Maj. | 512 Maj. | 27000 Maj. |
| Lucida Arietis | 2 | 42 | 4 32 | 59 T. | 2744 Min. | | 343000 Min. |
| | | | | 660 H. | 287496000 Maj. | 512 Maj. | 27000 Maj. |
| Cingul. Orionis 1. | 2 | 42 | 4 32 | 59 T. | 2744 Min. | | 343000 Min. |
| | | | | 660 H. | 287496000 Maj. | 512 Maj. | 27000 Maj. |
| Secunda Arietis | 3 | 35 | 3 47 | 49 T. | 4913 Min. | | 614125 Min. |
| | | | | 546 H. | 162771336 Maj. | 729 Maj. | 15285 Maj. |
| Tertia Arietis | 4 | 30 | 3 15 | 42 T. | 8000 Min. | | 100000 Min. |
| | | | | 468 H. | 102503232 Maj. | 1728 Maj. | 9526 Maj. |
| Eridani prima | 4 | 30 | 3 15 | 42 T. | 8000 Min. | | 100000 Min. |
| | | | | 468 H. | 102503232 Maj. | 1728 Maj. | 9526 Maj. |
| Sub axillâ Orionis quinta | 5 | 24 | 2 36 | 34 T. | 15625 Min. | | 1953125 Min. |
| | | | | 378 H. | 54010152 Maj. | 3375 Maj. | 5074 Maj. |
| Sub axillâ Orionis quarta. | 6 | 18 | 1 56 | 26 T. | 35937 Min. | | 4492125 Min. |
| | | | | 282 H. | 22425768 Maj. | 6859 Maj. | 2106 Maj. |

TABLE 1b -- Hevelius's data (continued).

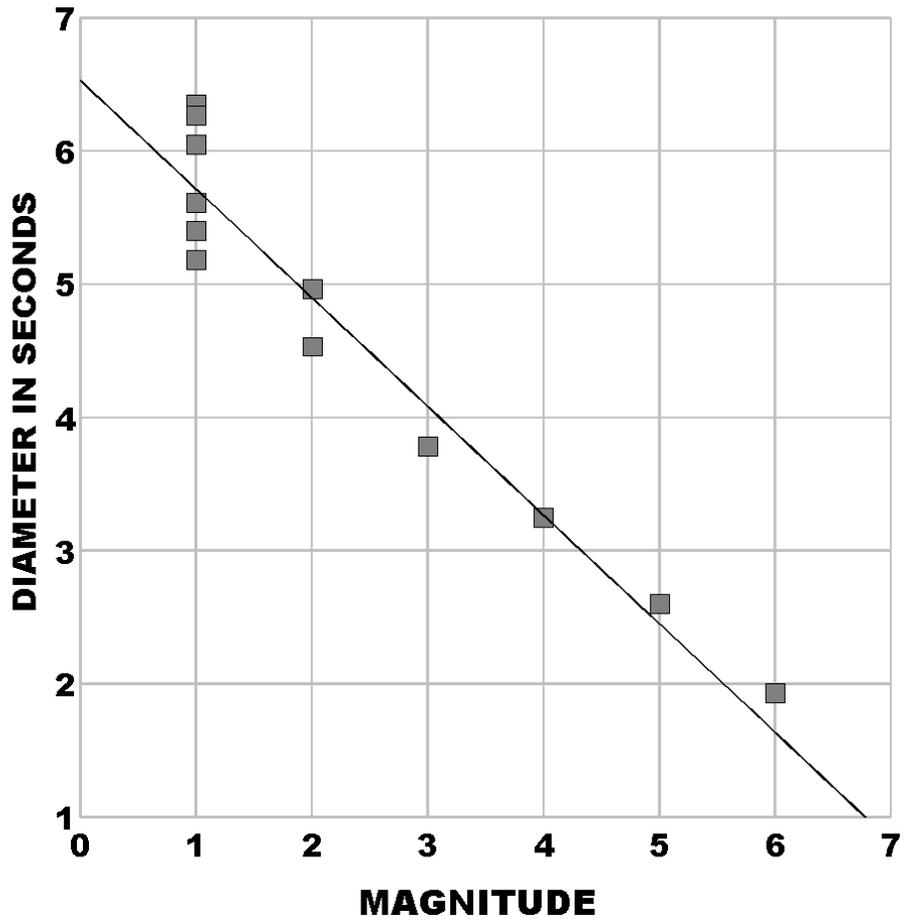

FIGURE 3 -- Plot of Hevelius's star image diameters vs. Hevelius's magnitudes.

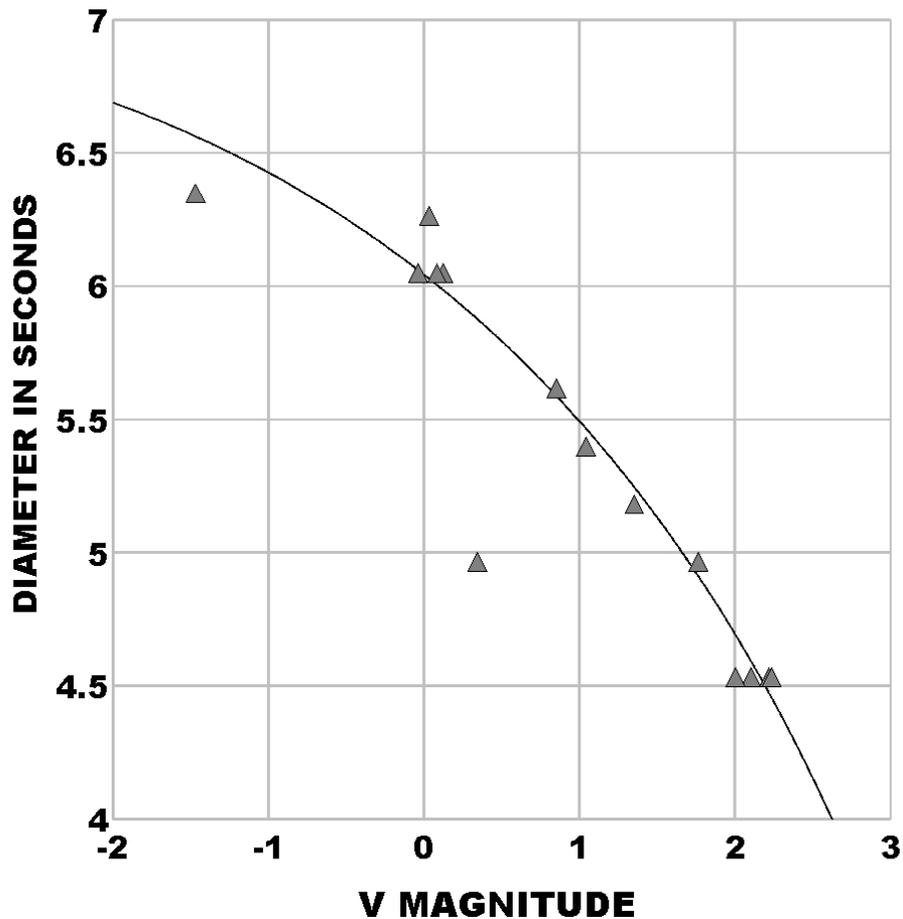

FIGURE 4 -- Plot of Hevelius's star image diameters vs. modern magnitudes from the SIMBAD database. The solid line is the APT model fit to Hevelius's data. Eleven of the stars hew closely to the APT model curve. Three stars break noticeably from the trend, most prominently Procyon (0.34, 4.97), but also Sirius (-1.47, 6.35) and Vega (0.03, 6.27). These deviations may be due to errors in the model. The deviations may be due to errors due Hevelius's measurements, or errors brought about by atmospheric effects or other changes in observing conditions – Hevelius provides limited information about his measurements, and in general he does not use modern methods for analyzing and reporting measurement errors, tending instead to make multiple measurements and select one of those measurements as his reported value (Buchwald 2006, pp. 585-586).  Finally, the deviations may be due to changes in magnitude from 1662 to today.  This plot is limited to stars listed by Hevelius as 1$^{st}$ and 2$^{nd}$ magnitudes, the brighter stars he measured.  The reasons for this are two-fold.  First, Hevelius includes a much smaller sample of higher-magnitude stars in his data; those few stars hew remarkably close to a line in the plot of Hevelius's diameters and magnitudes, suggesting that they may have been selected to illustrate a trend Hevelius believed to exist.  Second, Graney and Sipes (2009, pp. 98-104) note that the APT model is consistent with observations only over a limited range of brighter magnitudes, and for fainter stars needs to be refined with a second threshold to reflect the two types of receptor cells in the human eye.

| HEVELIUS | | | | SIMBAD | | | | |
|---|---|---|---|---|---|---|---|---|
| | | Apparent Diam. | | | | | | Apparent Diam. (sec) |
| Star Name | Mag | " | ''' | Name | ID | SAO | V Mag | |
| Sirius | 1 | 6 | 21 | Sirius | alf Cma | 151881 | -1.47 | 6.35 |
| Lucida Lyrae | 1 | 6 | 16 | Vega | alf Lyr | 67174 | 0.03 | 6.27 |
| Regel Orionis | 1 | 6 | 3 | Rigel | bet Ori | 131907 | 0.12 | 6.05 |
| Capella | 1 | 6 | 3 | Capella | alf Aur | 40186 | 0.08 | 6.05 |
| Arcturus | 1 | 6 | 3 | Arcturus | alf Boo | 100944 | -0.04 | 6.05 |
| Palilicium | 1 | 5 | 37 | Aldebaran | alf Tau | 94027 | 0.85 | 5.62 |
| Spica | 1 | 5 | 24 | Spica | alf Vir | 157923 | 1.04 | 5.40 |
| Regulus Leonis | 1 | 5 | 11 | Regulus | alf Leo | 98967 | 1.35 | 5.18 |
| Prima Caudae Ursae Majoris. | 2 | 4 | 58 | Alioth | eps Uma | 28553 | 1.76 | 4.97 |
| Procyon | 2 | 4 | 58 | Procyon | alf CMi | 115756 | 0.34 | 4.97 |
| Lucida Coronae | 2 | 4 | 32 | Alphecca | alf CrB | 83893 | 2.21 | 4.53 |
| Caput Serpentarii | 2 | 4 | 32 | Rasalhague | alf Oph | 102932 | 2.10 | 4.53 |
| Lucida Arietis | 2 | 4 | 32 | Hamal | alf Ari | 75151 | 2.00 | 4.53 |
| Cingul. Orionis I. | 2 | 4 | 32 | Mintaka | del Ori | 132220 | 2.23 | 4.53 |

TABLE 2 -- Hevelius's data with SIMBAD data.

(Graney 2007, Graney 2008, Graney and Sipes 2009), is not an an isolated or anachronistic event without context.  The context indeed predates Galileo and postdates Hevelius.  The following is a chronological listing of reports by visual astronomers, of stars as having measurable telescopic disks, usually with an explicit report that disk size decreases with magnitude:  Simon Marius (1614; Dreyer 1909); Galileo, observing notes of 1617 (*Le Opere di Galilei* III, Pt. 2, pp. 877, 880; Ondra 2004; Siebert 2005); Galileo, Letter to Ingoli of 1624 (Finocchiaro 1989, pp. 167-180); Galileo, *Dialogue* of 1632; Hevelius (1662); Flamsteed (1702); Cassini (see Halley 1720); Halley (1720); W. Herschel (1782); J. Herschel (1824); various 19$^{th}$ century astronomers working to reconcile the wave ("undulatory") theory of light's Airy Disk Radius, dependent only on aperture and wavelength, with the known phenomenon of stellar disk size decreasing with magnitude (Knott 1867; Hunt 1879; Glaisher 1886).[1]  This listing is simply what the author has found to date.  It is not meant to be comprehensive.  It is meant to provide the reader with a sense of which astronomers viewed stars as being disks.  The reports here range from simple mentions that stars seen through a telescope indeed do show disks (Flamsteed, Cassini) to statements that stars show disks whose size depends on magnitude (Marius, Halley, J. Herschel) to relative size measurements (Galileo, W. Herschel) to complete tables of measurements comparable to Hevelius's (Knott).

---

1   The answer they reached was the APT model:
> We think that the variation in the size of the spurious disk according to the brightness of the star may be explained by the circumstance that, according to the Undulatory Theory, the light fades away gradually from the central point outwards to the first dark ring, and that with the fainter stars it is only the central portion which is sufficiently bright to produce a sensible impression. Sir G. Airy has not given the diameter of the spurious disk, but that of the first dark ring, which is its *extreme* limit [Hunt 1879, pp. 152-153].

The second conclusion is that stellar photometry more sophisticated than guesswork or estimation was possible very early on in the history of astronomy.  It is unclear whether Hevelius or anyone else perceived this to be the case; clearly the concept never entered our historical record.  However, Hevelius's data exists, and given the number of astronomers who noticed stellar disks and their variation with magnitude, it is likely that similar historical data exists besides just Hevelius's (and Knott's).  Besides being of general historical interest, such photometric data may be useful to modern astronomers.  For example, Hevelius's data raises questions about Procyon, which deviates noticeably from the trend in Figure 4 and which Hevelius records as $2^{nd}$ magnitude while SIMBAD says is $0^{th}$ magnitude.  This could be simply a rather gross error on Hevelius's part, but finding another set of historical data similar to Hevelius's might be of interest to modern astronomers (see Gatewood and Han 2006).  Data from Hevelius has already proven useful to astronomers studying variable stars such as CK Vul (Shara, Moffatt, Webbink 1985). Historical astronomical data are becoming easier to find with the passage of time, as deposits of rare historical astronomical documents are made available electronically.  The Hevelius data used in this paper were found in a copy of *Mercurius* held by the Digital Library of the Gdańsk Library of the Polish Academy of Sciences and available online in DJVU format.  The author discovered Hevelius's data through a secondary source, a 19th century astronomy history book (Grant 1852, p. 545) which itself was available online via Google Books, and which was discovered through searches on terms such as "star" and "disk".  Not long ago obtaining a copy of *Mercurius* would have been a challenge.  Today it is available on any astronomer's desktop for free.

In summary, Johannes Hevelius's 1662 table of observations of stellar magnitudes and diameters published in his *Mercurius in Sole Visus Gedani* are consistent with the Airy pattern with threshold (APT) model used previously in this journal to explain the observations of Galileo Galilei.  The data thus helps to confirm the validity of the model, and to provide historical context for understanding early telescopic observations of the stars.  Moreover, Hevelius's data also suggest that telescopic astronomers in the 17th century could obtain photometric data of some precision -- data which may be useful to modern researchers.  Such data are becoming easier to find and are helping to illuminate a part of astronomy's history overlooked by historians.

ACKNOWLEDGEMENTS

I thank an anonymous referee for helpful comments, CDS for the SIMBAD database, and the Gdańsk Library of the Polish Academy of Sciences for making *Mercurius* available to all.Graney – Hevelius pg. 19

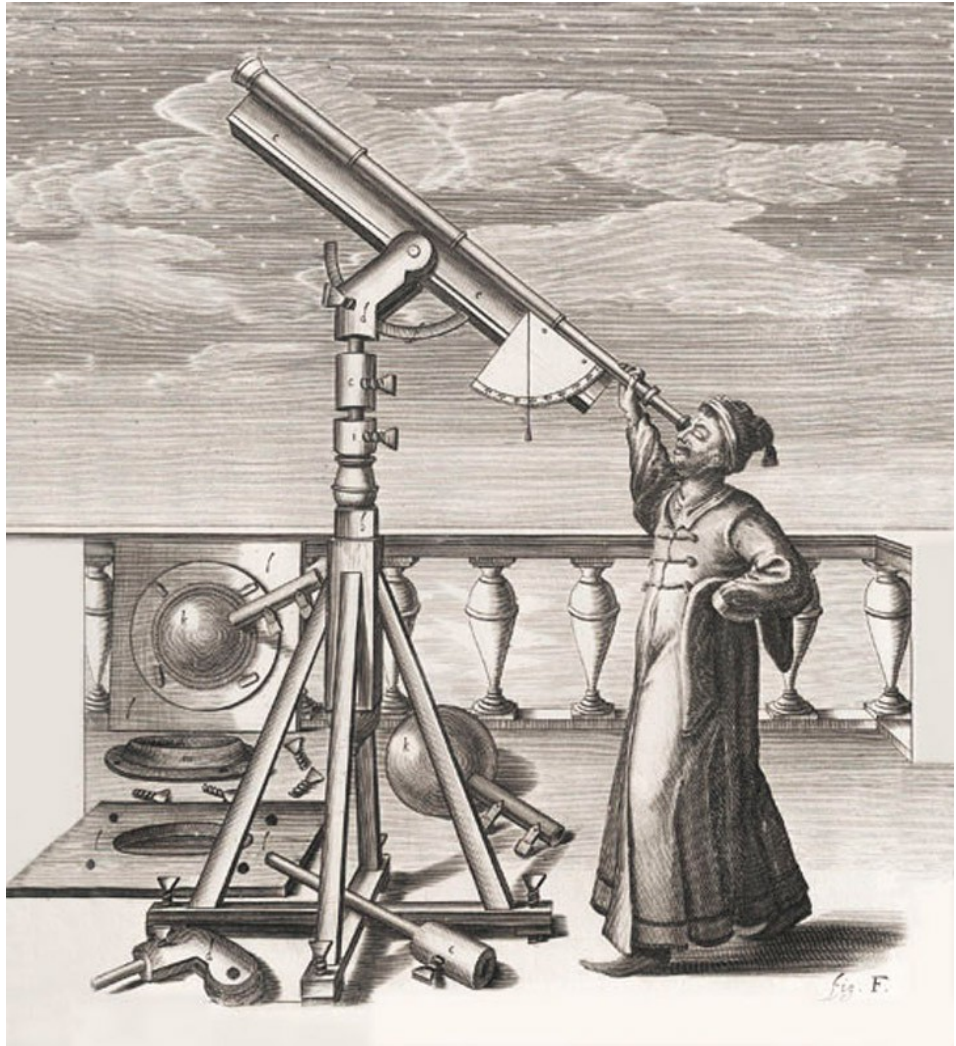

**FIGURE 5 -- The APT model was fit to the Hevelius data by adjusting the model's detection threshold and Airy Disk Radius. The Airy Disk Radius for the fit was 3.6", corresponding to an aperture of 38 mm using a wavelength of 550 nm. This value is consistent with the sort of instrument Hevelius would use, such as is depicted in the image of Hevelius shown here.**